\documentstyle[12pt,aasms4]{article}

\lefthead{Bekki Kenji}
\righthead{Unequal-mass mergers}

\begin{document}
\title{Unequal-mass galaxy mergers and the creation of cluster S0 galaxies}

\author{Kenji Bekki} 
\affil{Astronomical Institute, Tohoku University, Sendai, 980-8578, Japan \\
email address: bekki@astroa.astr.tohoku.ac.jp}

\begin{abstract}
 It is a longstanding and remarkable problem when and how red S0 galaxies
 were formed in clusters of galaxies.
 We here propose that the major mechanism for the S0 creation 
 is galaxy merging between two spirals with unequal mass.
 Our numerical simulations demonstrate that galaxy merging 
 exhausts a large  amount of interstellar medium of two gas-rich
 spirals owing to the moderately enhanced star formation, and
 subsequently transforms the two into
 one gas-poor S0 galaxy with structure and kinematics strikingly
 similar to the observed ones.
 This secondary S0 formation with enhanced star formation 
 explains
 a smaller fraction of S0 population
 recently observed in some distant clusters
 of galaxies. 
 Unequal-mass galaxy mergers  
 thus provide an evolutionary link between a larger number   
 of blue spirals  observed in intermediate redshift clusters 
 and red S0s  prevalent in the present-day ones. 

\end{abstract}


\section{Introduction}
 Since S0 galaxies were hypothetically introduced as a transition type
 between early-type elliptical galaxies and late-type spiral ones (Hubble 1936),
 the origin of S0s has been extensively explored both by observational
 studies and by theoretical ones (van den Bergh 1976, 1990; Bothun 1982).
 In particular, the origin of S0s in clusters of galaxies has been 
 discussed as to whether S0s are  products of 
 some particular protogalactic initial conditions
 or by-products of some secondary events.
 The previously proposed theoretical models for S0 formation are 
 high gaseous pressure of clusters environments (Evrard 1991),
 ram pressure (Farouki \& Shapiro 1980),  tidal encounters (Iche 1985),
 tidal compression by gravitational field of clusters (Byrd \& Valtonen 1990), 
 and tidal truncation  of gas replenishment (Larson, Tinsley, \& Caldwell 1980). 
 Although the observed  morphological,
 structural, kinematical, and spectrophotometric properties of 
 the present-day S0s have actually provided valuable constraints
 on the S0 formation (Dressler 1980; Kormendy \& Illingworth 1982; Fisher, Franx, \& Illingworth 1996;
 Fisher 1997),
 it remains unclear which theoretical model dominates
 the formation of S0s.

 Recent discoveries of a smaller fraction of S0 population 
 in distant clusters of galaxies at  the redshift of  $0.3 \sim 0.5$
 have shed  new light on the S0 formation (Dressler et al. 1997; Couch et al. 1998;
 van Dokkum et al. 1998; But see Andreon 1998); Morphological transformation
 into  S0s  is ongoing  in intermediate redshift clusters.
 These observations, combined with the observed
 larger fraction of  blue late-type spirals 
 in some  distant clusters (Butcher \& Oemler 1978; Lavery \& Henry 1994; Couch et al. 1994, 1998;
 Lubin et al. 1998), 
 suggest that
 some mechanisms drive dramatic  evolution  from blue spirals into red S0s in the course
 of cluster evolution.
 Then, what is the major mechanism that causes both the morphological transformation
 and the spectrophotometric changes ?
 We here consider that galaxy merging between two spirals with $unequal$ $mass$
 is one of promising candidates for the S0 formation.
 This unequal-mass merger model has not been extensively investigated  in previous studies.
 In this Letter,
 we use gas-dynamical numerical simulations with a plausible star formation model
 and thereby demonstrate how successfully the present merger model  can reproduce
 the observed properties of  S0s.
 What should be stressed here is  that the observed S0 evolution indicates only $some$ fraction of cluster S0s
 are formed by secondary events like unequal-mass mergers in intermediate redshift clusters,
 primarily because the considerably tight color-magnitude relation of S0s located in the cores of distant clusters
 (Ellis et al. 1997; Stanford et al. 1998) suggests  coevality of $some$ S0s and their formation at high redshift.
 Accordingly, we emphasize that the present study  provides  only a  new promising 
 candidate  for the  formation of some S0s in distant clusters of galaxies.

\section{Model}
  We construct models of galaxy mergers between two bulgeless gas-rich disks
  embedded in  massive dark
  matter halos by using the Fall-Efstathiou  model (Fall \& Efstathiou 1980).
  Initial mass-ratio of dark matter halo
  to disk stars to disk gas is 20:4:1 for the two  disks.
  The mass ratio of the larger disk (referred to as the `primary') to the smaller one
  (the `secondary') in a merger, which is represented by $m_{2}$,
  is considered to be a critical determinant
  for S0 formation in the present study.
  We mainly describe the results of the model with $m_{2}=0.3$,
  which displays  a typical behavior of S0 formation.
  The S0-like merger remnants with central spheroidal components
  are formed in the model with $m_{2}=0.3 \sim 0.5$.  
  Furthermore the results of two reference models with $m_{2}=0.1$  (`minor merger')
  and 1.0 (`major merger') are  briefly described for comparison. 
The disk mass ($M_{\rm d}$) is 
6.0 $\times$ $10^{10}$ $ \rm M_{\odot}$ for the primary
in all merger models of the present study.
The  exponential disk scale length  
and the maximum rotational velocity for disks are 
$3.5 {(M_{\rm d}/6.0 \times 10^{10}  \rm M_{\odot})}^{1/2} $kpc
and $ 220  {(M_{\rm d}/6.0 \times 10^{10} \rm M_{\odot})}^{1/4}$ km/s,
respectively.
These scaling relations are
 adopted so that
 both the Freeman law and the Tully-Fisher relation with
 a constant mass-to-light ratio can be satisfied for structure and kinematics of the two disks. 
 Parameter values for disk structure and kinematics for the model with $m_{2}=0.3$ are as follows.
 The size and mass of  a disk are set to be 
 17.5 (9.6) kpc and 6.0 (1.8)$\times 10^{10} \rm M_{\odot}$, respectively,
 for the primary (the secondary).
 The scale length and the scale height of an initial exponential disk is
 3.5 (1.9) kpc and 0.7 (0.38) kpc, respectively,  for the primary (the secondary). 
 The rotational curve of the primary (the secondary)
 becomes flat at 6.1 (3.4) kpc with the maximum velocity of 220 (163) km/s.
 The Toomre stability parameter (Binney \& Tremaine 1987) for initial disks is set to be 1.2.

Collisional and dissipative nature of disk interstellar gas is represented by
discrete gas (Schwarz 1981).
 Star formation in gas clouds during galaxy merging is modeled by converting gas particles
 to stellar ones according to the Schmidt law with  the exponent
 of 2.0 (Kennicutt 1989).
 Stellar components that are originally gaseous  ones are referred to as new stars. 
 The present numerical results  do not depend so strongly on the exponent
 in the Schmidt law.
 The thermal and dynamical effects of supernovae feedback  
 are not included, however, such effects probably modify the present 
 results only slightly.
 The intrinsic spin vector of the primary is exactly parallel with $z$ axis
 whereas that of  the secondary is tilted by ${45}^{\circ}$ from $z$ axis.
 The orbital plane of the merger is tilted by ${30}^{\circ}$  from $xy$ plane.
The orbit of a galaxy merger is assumed to be parabolic
and its pericenter distance and initial separation of the two disks  
are set to be 17.5 kpc and 140 kpc, respectively.
In this orbital configuration of galaxy merging,
the two disks merge with each other to form an early type galaxy  (E/S0, depending on $m_{2}$) 
within a few  Gyr.
Although global S0 morphology in the present merger model depends on initial orbital configurations of galaxy
merging, this dependence will be described in our future paper (Bekki 1998).
All the simulations have been carried out on the GRAPE board (Sugimoto et al. 1990)
with the
gravitational softening length of 0.53 kpc.

\section{Results}
Figure 1 describes the morphological evolution of a gas-rich galaxy merger with $m_{2}$=0.3.
As galaxy merging proceeds, the primary forms tidal tails and the secondary 
gradually sinks toward the center of the primary.
The primary suffers heavily from vertical dynamical heating of galaxy merging 
and gradually loses its late-type morphology.
The transient feature of the simulated merger is remarkably similar to 
some blue interacting/merging
galaxies with disturbed morphology observed in distant clusters (Lavery \& Henry 1994;
Couch et al. 1994, 1998) and consistent with previous numerical results of Mihos (1995).
As is shown in Figure 2,
after a substantial amount of disk gas is converted into stellar components
by star formation,
the two unequal mass spirals are finally transformed into one S0 galaxy that has a  flattened 
oblate spheroid, an outer diffuse  stellar envelope (or diffuse disk-like component
with large scale-height),
and a central thin stellar bar composed mainly of
new stars.
This result suggests that the outer stellar envelope frequently observed in S0s is  
originally the outer part of the primary (old) stellar disk that is dynamically 
heated up and thickened during galaxy
merging, and furthermore that the central luminous part in a S0 results from
efficient inward mass-transfer in a galaxy merger.
Figure 2 furthermore shows that 
owing to the  violent relaxation combined with  gaseous dissipation during merging,
the merger remnant  shows both the central de Vauclouleurs  density profile and  
a moderate amount of global rotation.
The outer part of the merger remnant (the radius larger than 10 kpc) still seems to
show shallower exponential profile, which is similar to the luminosity profile 
characteristic of S0s.
Applying the two-components fitting (Kormendy 1977)
to the merger remnant
(that is, assuming that the density profile of the merger remnant is the combination of 
the de Vauclouleurs profile and the exponential one),
we found that the bulge-to-disk-ratio is  2.16.
Radial mass distribution,
a rotationally flattened spheroid,  and a diffuse  stellar envelope
(or a thick stellar disk-like component) in the simulated S0 are well consistent  
with the observed S0 properties (Kormendy \& Illingworth 1982; Burstein 1979; Fisher 1997).

Figure 3 shows that two times moderate bursts of star formation with the peak rate of 18.4 $\rm M_{\odot}$/yr,
which are induced by tidal interaction
and merging,  exhaust 75 \% of  disk gas corresponding to  
9.0 $\times$ $10^{9}$ $ \rm M_{\odot}$ during galaxy merging.
The first  peak of star formation rate (18.4 $\rm M_{\odot}$/yr)
is due to enhanced cloud-cloud collisions
and the subsequent rapid increase of gas density in the first encounter between
two gas-rich disks.
The second peak (8.2 $\rm M_{\odot}$/yr), on the other hand, results from 
the final merging of the gas-rich secondary.
The mean star formation rate for $0.0 \leq T < 0.7$ Gyr (before merging),
$0.7 \leq T < 2.3$ Gyr (during interaction/merging),
and $2.3 \leq T < 3.9$ Gyr (after merging)
are 2.9 $ \rm M_{\odot}/ \rm yr$, 5.6  $M_{\odot}/ \rm yr$,
and 0.55 $ \rm M_{\odot}/ \rm yr$, respectively.
Mean star formation rate is  drastically reduced to 0.55 $\rm M_{\odot}$/yr after 
the completion of galaxy merging.
These secondary starbursts followed by the nearly truncated star formation 
are in good agreement with S0 star formation histories  inferred from 
the scatter in the color-magnitude relation of
distant cluster S0s (van Dokkum et al 1998).
The gas located initially in the considerably outer part of the primary 
is tidally stripped during merging to form diffuse  gaseous halo surrounding the merger
remnant.
This tidal stripping 
also reduces the total amount of  gas finally within the remnant.
These results suggest that the origins of the observed deficiency of gas in S0s
can be closely associated with rapid gas consumption by starbursts 
and tidal stripping of gas in galaxy mergers.

A galaxy merger roughly 1 Gyr after the completion of the secondary  starbursts
can  be identified
as a  galaxy with the so-called `E+A' spectra (Dressler \& Gunn 1992),
which are characterized by 
strong Balmer absorption lines and no significant [OII] emission.
Furthermore the merger remnant that
has younger stellar populations formed in the isolated disk
evolution and the galaxy merging 
naturally shows bluer colors until the hot and blue massive stars die out.
In addition, the  gas-poor remnant
may well become a red and passively evolving
S0 a few Gyr after the completion of galaxy merging. 
Galaxy merging with starbursts accordingly provides  an  evolutionary link between blue spirals,
mergers with `E+A' spectra, S0s with bluer colors, 
and red bona fide S0s,
all of which are actually observed in clusters 
environments with different redshifts (Butcher \& Oemler 1978;
Lavery \& Henry 1994; Couch et al. 1994, 1998; Dressler et al. 1997; van Dokkum et al. 1998).
What we should emphasize here is that S0 morphology of merger remnants in the present model  
is not  a representative of `E+A' galaxies in distant clusters of galaxies (Couch et al. 1998);
Only some fraction of `E+A' galaxies observed in distant clusters probably become red S0s
through unequal-mass galaxy merging with starbursts.

 What  we should emphasize furthermore is that 
 the merger remnant in the model with $m_{2} =0.3$ 
 is more morphologically similar to S0s
 than those in the minor merger model with $m_{2} =0.1$ and the major one with
 $m_{2} =1.0$.
 The primary is  discernibly  thickened  owing to dynamical heating
 of galaxy merging and does not develop
 a large spheroidal component in the minor merger model whereas
 the primary  is completely destroyed  to form a spheroid
 looking more like an elliptical galaxy in the major merger one.
 Dependences  of structural, kinematical, and morphological evolution 
 on $m_{2}$  are described in detail  by  Bekki (1998).
 These results imply that the merger mass ratio ($m_{2}$)
 is required to be within a certain range for S0 formation.
Major galaxy merging ($m_{2} \sim 1.0$) has been already
confirmed to reproduce fundamental structural, kinematical, and photometric
properties of elliptical galaxies (Barnes \& Hernquist 1992; Schweizer \& Seitzer 1992).
The present merger model of S0 formation thus 
suggests that the mass ratio
of two merger progenitor disks ($m_{2}$) is a critical determinant for
generating differences in physical properties between ellipticals
and S0s.

\section{Conclusions}

We conclude that  galaxy merging between two gas-rich spirals with unequal mass inevitably
exhausts a substantial amount of disk gas owing to the induced secondary starbursts 
and subsequently transforms the two into one gas-poor S0 galaxy with morphology,
structure, and kinematics strikingly similar to the observed ones in
S0 galaxies.
Galaxy merging with starbursts has some advantages in changing drastically structure,
kinematics, star formation history, and gas content of galaxies, and thus
provides not only an evolutionary link
between blue spirals and red S0s but also a critical determinant for 
the differences of S0s and ellipticals.
Furthermore galaxy merging with secondary starbursts  provides a
clue to the physical connections between `Butcher-Oemler'  and 'E+A' galaxies in intermediate redshift 
clusters and red S0 galaxies in the present-day  ones.

Counter-rotating components frequently observed in S0s (Bertola, Buson, \& Zeilinger 1992)
and small color differences
between S0 bulges and disks (Peleteir \& Balcells 1996)
probably support this secondary S0 formation.
Furthermore, the existence of field S0s and a large fraction
of `E+A' interacting/merging
galaxies $-$ promising progenitor of ellipticals 
and S0s $-$ in field environments (Zabludoff et al. 1996) 
favor the mechanism that is not necessarily associated
with cluster evolution  and accordingly strengthen the viability of the merger model
of S0 formation.
Even galaxy merging with the widely spread merging epoch has been demonstrated to keep 
the tight color-magnitude relation of E/S0 galaxies (Shioya \& Bekki 1998),
which suggests that the nature of stellar populations in S0s can be reproduced equally by galaxy
mergers.

The merger model, however,
has not yet addressed  
radial dependences of galactic morphology and colors observed in nearby and distant
clusters (Dressler 1980; Abraham et al. 1996; van Dokkum el al. 1998),
each of which is an important observation that should be explained by any theories.
Future high-resolution numerical studies including both large-scale cluster dynamics
and spectrophotometric evolution of individual galaxies will confirm whether or not
the merger scenario can give  a self-consistent and convincing explanation for
these observational results.
Furthermore, the present merger model has not yet provided a reasonable explanation
for variously different morphology of S0s such as S0s  with thin stellar disks (e.g., NGC 4111), 
S0s with boxy spheroid (e.g., NGC 128), polar ring S0s (e.g. NGC  2685).
Our future studies will confirm whether or not the observed structural and kinematical
diversity of S0s with different luminosity (van den Bergh 1990) 
can be reproduced successfully in galaxy mergers.
The presented merger model is only one of promising candidates of S0 formation.
Accordingly, forthcoming observational studies  on  stellar populations of S0s
and detailed radial distribution of galactic morphology in more distant clusters of galaxies
will assess the validity of the present merger scenario of S0 formation
and determine which theoretical mechanism  dominates the  S0 formation.

\acknowledgments

We are  grateful to the anonymous referee for valuable comments, which contribute to improve
the present paper.
K.B thanks to the Japan Society for Promotion of Science (JSPS)
Research Fellowships for Young Scientist.


\figcaption{
 Morphological evolution of a galaxy merger between two gas-rich bulgeless
 spirals with star formation projected onto $xy$ plane (upper)
 and $xz$ one (lower)
 in the model with $m_{2}=0.3$.
 Stars, gas, and new stars are plotted 
 in the same panel.
 We here do not display halo components in order to  show more clearly 
 the morphological evolution of disk components.
 Each frame measures 142 kpc, and the time, indicated in the upper left
 corner of each panel,  is in units of Gyr.}

\figcaption{
 The top and middle two panels show face-on view (left) and edge-on one (right)
 for stars and new stars, respectively,
 in  the merger remnant of the model with $m_{2}=0.3$
 (Each frame measures 35.5 kpc). 
 The bottom left and right panel 
 show the projected major-axis mass distribution (a solid line  with open squares)
 and the radial velocity profile
 along the major axis 
 (a solid line  with open squares), respectively.
 For comparison, initial exponential mass distribution 
 of the primary is also indicated by a solid  line 
 in the  bottom left panel.
 A solid horizontal line and dotted one in the bottom  right panel
 represent initial value of maximum rotational velocity
 of the primary and that of the secondary, respectively.
 Note that the merger remnant shows both clear deviation
 from initial exponential density profile 
 and  a moderate amount of 
 global  rotation.
 This merger remnant is more similar to S0s with no remarkable thin stellar disks
 such as NGC 3245 and 4684 than those with thin extended stellar disks such
 as NGC 4111 and 4710 (See the Hubble Atlas of Galaxies, Sandage 1961).
This  merger remnant
can be morphologically classified either as a  $\rm S0_{1}$ or as a SB0.}

\figcaption{
 The time evolution
 of star formation rate (upper) and that of the fractional gas mass (lower)
 for the model with  $m_{2}=0.3$ (solid lines with triangles).
 For comparison, the results for two reference models with  $m_{2}=0.1$ (solid lines)
 and 1.0 (dotted  ones) are also given.
 Owing to a large initial gas mass fraction (0.2),
 the star formation rate in disks is rather high for the time $T$ less than 0.5
 Gyr, during which two disk can be considered to show isolated evolution.
 It should be emphasized that the magnitude of starbursts in the model with  $m_{2}=0.3$
 is rather moderate compared with that in the major merger model ($m_{2}=1.0$).}

\end{document}